\documentclass[conference, 10pt]{IEEEtran}
\IEEEoverridecommandlockouts
\usepackage{cite}
\usepackage{amsmath,amssymb,amsfonts}
\usepackage{algorithmic}
\usepackage{graphicx}
\usepackage{textcomp}
\usepackage{xcolor}
\def\BibTeX{{\rm B\kern-.05em{\sc i\kern-.025em b}\kern-.08em
    T\kern-.1667em\lower.7ex\hbox{E}\kern-.125emX}}
    
\usepackage{todonotes}    
\usepackage[absolute,overlay]{textpos}

\usepackage{etoolbox}
\DeclareListParser{\doslashlist}{/}
\newcounter{ndnNameComponentCounter}%
\newcommand{\ndnName}[1]{{%
  \setcounter{ndnNameComponentCounter}{0}%
  \renewcommand{\do}[1]{{%
    \ifnumgreater{\value{ndnNameComponentCounter}}{0}{\allowbreak}{}%
    \ifnumodd{\value{ndnNameComponentCounter}}{}{}%
    /##1}%
    \stepcounter{ndnNameComponentCounter}}%
``{\fontfamily{Arial}\selectfont\doslashlist{#1}}''%
}}

\usepackage{amssymb}

\begin{document}

\begin{textblock*}{20cm}(1cm,1cm) 
{\color{red}{This is the authors' version of the work. The final version of this paper will appear at LCN 2022.}}
\end{textblock*}

\title{Vehicle-to-Vehicle Charging Coordination over Information Centric Networking}

\author{\IEEEauthorblockN{Robert Thompson}
\IEEEauthorblockA{{Department of Computer Science} \\
{Tennessee Tech University}\\
Email: rjthompson42@tntech.edu}
\and
\IEEEauthorblockN{Muhammad Ismail}
\IEEEauthorblockA{{Department of Computer Science} \\
{Tennessee Tech University}\\
Email: mismail@tntech.edu}
\and
\IEEEauthorblockN{Susmit Shannigrahi}
\IEEEauthorblockA{{Department of Computer Science} \\
{Tennessee Tech University}\\
Email: sshannigrahi@tntech.edu}
}

\maketitle

\begin{abstract}

Cities around the world are increasingly promoting electric vehicles (EV) to reduce and ultimately eliminate greenhouse gas emissions. For example, the city of San Francisco  aims to increase the number of EVs from tens of thousands to over quarter of a million by 2025. This huge number of EVs will put unprecedented stress on the power grid. To efficiently serve the increased charging load, these EVs need to be charged in a coordinated fashion. One promising coordination strategy is vehicle-to-vehicle (V2V) charging coordination, enabling EVs to sell their surplus energy in an ad-hoc, peer to peer manner. 

Enabling V2V charging coordination requires new communication network protocols that can facilitate such charging coordination in a peer-to-peer fashion. This paper introduces an Information Centric Networking (ICN)-based protocol to support ad-hoc V2V charging coordination (V2V-CC). Our evaluations demonstrate that V2V-CC can provide added flexibility, fault tolerance, and reduced communication latency than a conventional centralized cloud based approach. We show that V2V-CC can achieve a 93\% reduction in protocol completion time compared to a conventional approach. We also show that V2V-CC also works well under extreme packet loss, making it ideal for V2V charging coordination.

\end{abstract}

\section{Introduction}

With electric vehicle (EV) 
adoption on the rise 
along with estimates of their 
increasing integration into smart cities, charging demands of these vehicles will also increase. 
This new increasing electric 
load 
can be supported only by the largest and most modern power grids. 
Previous studies have shown that even a $10\%$ increase in EV load concentration can significantly stress the power grid and result in blackouts \cite{liu2011survey}. To compound this issue, 
EV charging tends to occur in bursts, where many 
EV owners all 
start and complete their charging around the same time, placing an extreme load on the grid 
at one time instance. 

Previous work has proposed two solutions. The first approach is to upgrade the power grid capacity in order to be able to accommodate the increasing penetration level of EVs. In addition to the expected high investment cost, the upgraded capacity might only be needed for a short duration throughout the day to serve the peak EV charging demands. A more appealing solution is to coordinate the charging requests of EVs either temporally for parked EVs or both spatially and temporally for mobile EVs \cite{liu2011survey} in order not to stress the power grid. 
To better serve the expected charging load, we need not only stationary charging stations but also 
vehicle-to-vehicle (V2V) charging mechanisms that allow one 
EV to sell 
its surplus energy to other 
EVs. Adhoc V2V charging 
not only reduces the load on the grid infrastructure but also allows the users to avoid bottlenecks imposed by the power grid technical limits during 
high demand intervals.

To support efficient V2V charging coordination, rapid message exchanges among EVs are required. This message exchange protocol should offer high flexibility, scalability, and low communication latency while serving \textit{mobile} EVs. Unfortunately, the conventional IP-based centralized  protocols cannot offer such desirable features when mobility is introduced. Hence, 
in this 
paper, we introduce 
V2V-CC, a communication protocol for V2V charging coordination based on Information Centric Networking (ICN). This is the first work, to our knowledge, to propose a V2V charging coordination protocol using ICN. Utilizing ICN for direct V2V communication 
offers several benefits. Specifically, we show that V2V charging coordination over ICN can happen much faster compared to an IP-based centralized controller. In addition, 
the consumer 
enjoys more control over the 
seller selection process.

\section{Background}

\subsection{Electric Vehicle Charging Coordination}

Three modes can be distinguished for EV charging coordination, namely, grid-to-vehicle (G2V), vehicle-to-grid (V2G)\cite{das2020charging}, and V2V \cite{mou2019vehicle}.
For this work, we only 
focus on the V2V charging coordination where EVs exchange energy in an adhoc manner among sellers and buyers via bidirectional chargers without the need to go through the power grid \cite{mahure2020bidirectional}. When EVs are stationary at smart parking lots, 
only 
temporal coordination of charging requests is needed. 
In the context of this work, we need to coordinate the charging requests of 
\textit{mobile} EVs both on spatial and temporal dimensions so that we can reach consensus between buyers and seller on where and when to exchange energy. To do so, we need to gather information from the vehicles including the amount of charging request from demanding EVs, the amount of surplus energy from supplying EVs, and state-of-charge (SoC) of the EVs. 
Another very important piece of 
information is the range anxiety for both demanding and supplying EVs. The range anxiety reflects how far the EV can go to charge or discharge given the current battery state of charge (SoC) and the relevant distance. After all, the spatial coordinated decision should ensure that the EV battery does not get depleted before it reaches its charging point. The outcome of the charging and discharging coordination strategy specifies both the time instant to charge (for demanding EV) or discharge (for supplying EV) and the location where such a transaction will take place.

\subsection{Named Data Networking}

Named Data Networking (NDN) is a clean-slate redesign of the Internet and its networking protocols as an implementation of ICN. For the rest of the paper, we refer to ICN and NDN interchangeably. In NDN, a consumer can construct Interest packets with a given name in order to request a Data packets with the same name that is hosted by a producer. Rather than the end-points being identified by something like an IP address, the data itself is identified by the name. NDN is also a pull-based model where consumers ``pull" the data, as opposed to pushed-based models like IP where the servers ``push" the data to the destination. Lastly, since there are no named locations, any device and network can utilize application defined namespaces. In our example, an EV can simply broadcast a discovery Interest - any number of EVs providing charge may respond to the request. Under normal NDN operation, the first answer will reach the requester. Depending on how many responses the requester needs to make a decision, it can send multiple requests into the network.  Once an Interest packet has reached a node with the Data  (either a node that has the data cached or the producer), a Data packet with the same name as the Interest is sent back through the network.
The returning data packets are cached at each hop and can satisfy future requests with the same name. Since a location based addressing scheme is not required and ad hoc networks can be more easily created, NDN also excels in lossy and highly mobile environments, making it ideal for vehicular communications required for V2V coordination.

\subsection{Vehicle Communication Using NDN}

While at the time of writing no other NDN-based V2V charging coordination protocols exist, significant research has been done for other NDN-based vehicular applications. \cite{xia2018improving} presents a V2V communication scheme that expands on vehicular adhoc networks (VANETs). Specifically, traffic data being transmitted over an NDN-enabled application stack between large numbers of vehicles in a highway environment was investigated. The authors not only showed an increase in time savings using NDN but also demonstrated how NDN's in-network caching can be used to 
enhance the communication service quality in such scenarios. Similarly, \cite{CHEN2014208} also presents findings leveraging NDN's design, implementing a popularly mechanism to ensure that more popular data has a longer lifetime within NDN's caching servers to leverage caching as much as possible. In addition, \cite{grassi2013vndn} provides unique insights on NDN vehicular communications, specifically addressing V2X by using NDN equipped vehicles and transmitted data to 
other vehicles as well as road side units.


 \section{Protocol Design}

V2V-CC allows 
EVs to communicate with each other in an adhoc manner. 
Based on an NDN protocol, there is no need for addressing individual 
EVs. The 
EV willing to provide charge (the seller) can simply announce a name prefix such as \ndnName{/FastCharging} indicating it is willingness to sell 
energy. A prefix can potentially be reserved specifically for charging coordination. Once a namespace is agreed upon (this is outside the scope of our paper), the producer and the consumer can start exchanging messages.

\begin{figure*}[!ht]
     \centering
     \includegraphics[width=1.45\columnwidth]{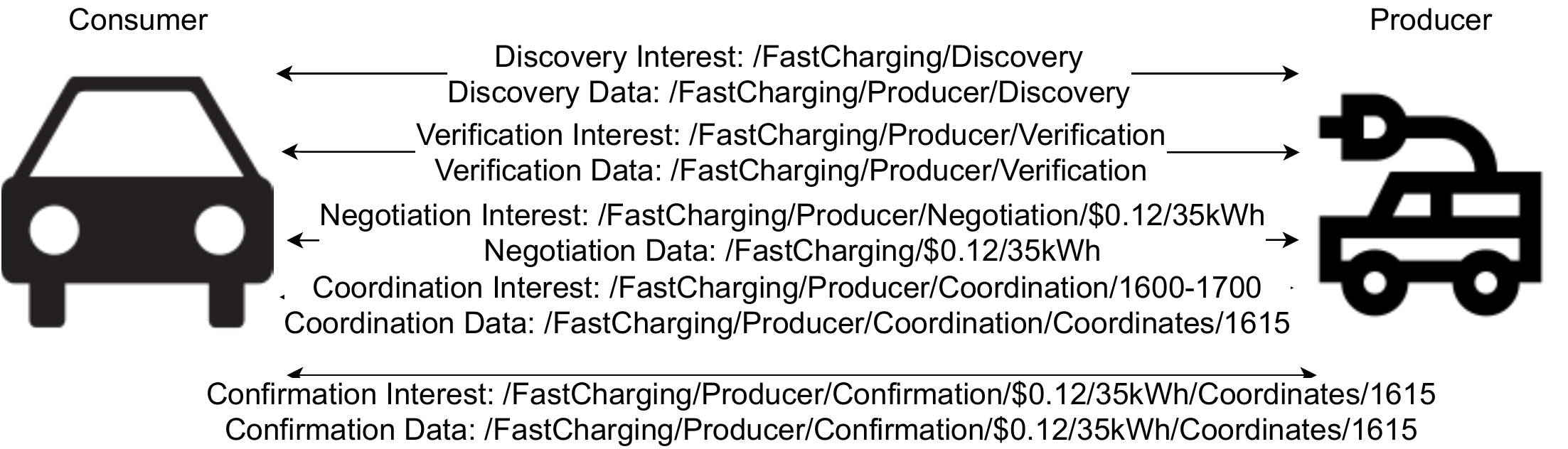}
     \caption{V2V-CC Example with each phase in detail}
     \label{fig:coordination}
 \end{figure*}

We logically break down the protocol into five phases - seller discovery, verification, negotiation, coordination, and confirmation phases. However, multiple phases can be combined for optimization. For example, discovery, verification, and (potentially) negotiation can all be done with one Interest/Data exchange. Similarly, coordination and confirmation phases can be combined together. In this paper, we keep these phases separate for showcasing the different phases of the proposed protocol. When phases are combined, the performance of the proposed protocol will further improve. Figure \ref{fig:coordination} shows an overview of V2V-CC. The following subsections discuss each of these phases in details.

\subsection{Seller Discovery Phase}
In order to begin 
V2V charging, a consumer needs to know 
which suppliers exist in the nearby area. 
To find such available suppliers, the consumer 
sends an initial discovery interest, which in its most basic form is constructed as: \ndnName{/FastCharging/Discovery/Timestamp}.  Any 
supplying EVs that want to 
sell energy can respond with a similarly constructed data packet: \ndnName{/FastCharging/Producer’s Identifier (PID)/Discovery}. Using NDN's stateful forwarding plane and native multicast, many 
suppliers can be reached at once. While only one data packet is needed to satisfy the interest, any other packets that are sent in reply can be cached in nearby NDN nodes, reducing the amount of hops it takes for subsequent requests to get a reply. 


A consumer is also able to insert a number of filters to their interest to find 
suppliers that meet certain requirements. A consumer may choose to limit their discovery search to a specific geographical area, 
or any number of other attributes in a single interest. For example, if a consumer wanted to discover 
supplying EVs that are within a $2$km radius, with an available time slot between 14:00 and 15:00 hours, with a cost per kWh no more than $\$0.10$, and no less than $25$ kWh of charge available, as well as a 
reputation of no less than $7$, the interest would be constructed as such: \ndnName{/FastCharging/Discovery/Current Location+2km2/0.10/25/7/1400/1500/Timestamp}. 
Once a single 
supplier is discovered, it is up to the consumer to decide whether to move on to the verification step of the protocol or if the consumer wants to discover more 
suppliers. Currently, the protocol defaults to searching for three 
suppliers before moving onto the verification phase, however, this number can be changed by the consumer at any time.  By the end of the discovery phase, a consumer will have a selection of 
suppliers with whom to continue on to the verification phase. 



\subsection{Verification Phase}
After the discovery of 
suppliers, the verification phase begins. This phase is designed to ensure that any information that arrives to a consumer is correct. By default, the consumer chooses the ``best" 
supplier with whom to communicate based 
on the information that was received during the discovery phase. 
Currently, the best 
supplier is determined by the lowest cost, breaking ties using the shortest distance and, if needed, reputation. These weights may be 
adjusted at any time by the consumer to match what is needed at the time of the coordination. For example, if a consumer needs charge urgently, they may choose to prioritize closest distance above all 
or if a consumer has a tight schedule, any 
supplier with an open time slot that best fits the consumers needs can be selected. Any information that was not given during the discovery phase can be requested during the verification phase as well.

Verification interests are constructed as such by default: \ndnName{/FastCharging/PID/Verification/Timestamp}. This naming construct allows for only the 
supplier with the PID in the name to respond to the request, essentially enabling point-to-point communication.  The 
supplier will respond with the corresponding data packet: \ndnName{/FastCharging/Verification/PID}. In this phase, the consumer will want to verify 
supplier's signature (enabled by default in NDN), verify additional data, and get new data that the consumer does not have. 



By the end of the verification phase, the consumer will have accurate and fresh data on one or more 
suppliers with whom the consumer can communicate during the next phase. This phase can also uncover some malicious 
suppliers that 
provide incorrect data during the discovery phase. The information requested or double checked during this phase can be as minimal or detailed as the consumer desires. 

\subsection{Negotiation Phase}

After 
all 
the data from the 
supplier is confirmed, the negotiation phase begins. This phase is the most variable of all 
phases due to its monetary nature. This phase can be as short as one interest and one data response or as many rounds of communication as it takes to come to some form of agreement. A negotiation interest looks like: \ndnName{/FastCharging/PID/Negotiation/Suggested Price/Suggested Charge Amount/Timestamp}.  Since the consumer has the base price and the amount of charge that the 
supplier is offering as a baseline from either the discovery phase, negotiation phase, or both, the consumer can ask for a lower price and the amount of charge it wants.  


A 
supplier's response at its core is sent as such: \ndnName{/FastCharging/PID/Negotiation/Counter Price/ Counter Charge Amount/Timestamp}. If the counter price and charge amount are the same as the consumers offer, the negotiation phase comes to an end. If it is not however, the consumer may respond with another constructed counter offer and the cycle continues. One way this can be limited is the addition of the “hard offer” flag, which then concludes the negotiation.

By the end of the 
verification phase, the consumer and 
suppliers will have a charge amount and a cost per kWh for that charge that has been agreed upon by both parties. Once the negotiation phase is complete and a charge amount is agreed upon, a 
supplier may choose to reserve that charge for a certain period of time while the next two phases complete.

\subsection{Coordination Phase}

After a suitable price has been negotiated between a consumer and producer, the coordination phase can begin. This phase has one main objective, to find 
time and place for the 
supplier to transfer 
energy to the consumer.

For V2V charging, both spatial and temporal coordination needs to happen. The base form of the interest in our protocol is : \ndnName{/FastCharging/PID/Coordination/Spatial/Temporal/Time Frame/Location} with time frame being the start of when the consumer is available and by when they wish to be done and the location 
being either empty or contain the suggested location for the 
energy to be exchanged. If the location is empty, it is up to the 
supplier to choose a location, which it will do selfishly. Since the 
supplier is mobile, it may choose a location that is close to other charging sessions that are just before or just after, a charging location it is familiar with like a parking lot, or simply its current location. If the location component is populated, the consumer is offering to meet at the listed location. The 
supplier can choose to meet there or counteroffer with a different location. In some cases, it may be advantageous for a 
supplier to travel to the consumer's location, especially if the consumer is offering to pay extra for the charge. 

If the 
supplier is mobile, it must take into account the travel time between two locations when responding to a consumer's temporal interest. Most of these calculations can be offloaded to an on board GPS, but we leave this for a future work.  After coordination is completed, both the consumer and 
supplier have decided on a time and place to exchange 
energy. All of the other information including the price and amount of charge has been established, so the final protocol phase in V2V-CC can be started.

\subsection{Confirmation Phase}
The purpose 
of the confirmation phase is to double check all of the information that was sent throughout the entire protocol and create a single point for logging the transaction if memory is limited. Since all of the data needs to be checked and not all charging coordination communication will be the same, the information in this phase will differ from confirmation to confirmation. The generic form for the name is \ndnName{/FastCharging/PID/Coordination/Negotiation/Timestamp} where coordination 
includes the the time 
and location as 
obtained from the coordination phase and negotiation consists of the amount of charge and the cost per kWH, from the phase with the same name. Any other information that is agreed upon during these two phases would also be included in their respective locations. Like with the verification phase, any other optional information can be inserted between the negotiation and timestamp information. 


As mentioned earlier, these five phases compose the basics of the protocol at hand, but they do not form a solid rule. Since 
V2V-CC is designed to be flexible, additional phases can be inserted or appended and phases can be combined or omitted entirely. 

\section{Simulation Setup}\label{simSetup}

To evaluate the feasibility of V2V-CC, we utilized ndnSIM \cite{mastorakis2015ndnsim}, a custom fork of NS-3 network simulator. Since 
V2V-CC is the first application of its kind, we designed and implemented custom 
suppliers and consumers. The 
suppliers are 
EVs that are selling 
energy and the consumers are 
EVs requesting charge. The consumer issued interests and utilizes the data returned from the 
suppliers. The 
suppliers are responsible for responding to consumer interests with data packets.  

Once the applications were developed, we put them in an environment that included node mobility. 
V2V-CC begins by creating 
``EVs" (nodes) that hold applications, mobility models, and connection parameters. Every simulation begins at timestep 0 and ends at a predefined time, set by the user.  


In the current setup, V2V-CC uses 
ad hoc Wi-Fi as the wireless medium for communication in the simulation since 5G or LTE models are not available with the current version of ndnSIM. We paid close attention to the parameters to create a realistic scenario according to 4G-LTE deployment requirements. 
More discussions are presented on this in Section VI.


Since 
V2V-CC needed to be 
evaluated alongside a benchmark of similarly implemented IP-based (central coordinator approach), we created another simulation and three more custom applications using the same version of NS-3. We developed three different applications, namely, a charge provider, central coordinator, and a client. The charge provider periodically informs the central coordinator of its current state. The central coordinator receives the provider information and supplies any connecting client with that data. Finally, the client requests the provider's information from the central coordinator and parses the data that it receives.

\begin{table}[!ht]
\caption{Simulation Parameters}
    \centering
    \begin{tabular}{|c|c|}
    \hline
         Number of Suppliers & $1-10$ \\
    \hline
         Number of Consumers & $1-21$ \\
    \hline
        Supplier to Consumer Ratio & $1$ Supplier to $3$ Consumers \\
    \hline
        Wi-Fi Bandwidth (links never fill) & $24$ Mbps \\
    \hline
        Number of Suppliers Discovered & $1$, $3$ \\
    \hline
        Timeout Wait Duration (ms) & $30$, $50$ \\
    \hline
       Artificial Loss & $0\%$, $20\%$ \\
    \hline
        Mobility Speed (mph) & $0, 10, 30, 50, 70$ \\
    \hline
        IP Bandwidth (links never fill) & $24$ Mbps \\
    \hline
        IP Connection Delay (ms) & $25, 50, 100$ \\
    \hline
        IP Error Rate (minimal effect on results) & $0.05\%$ \\
    \hline
        IP Number of Charge Suppliers & $1-3$ \\
    \hline
        IP Number of Clients & $1-30$ \\
    \hline
    \end{tabular}
    \label{tab:simulation_parameters}
\end{table}

\begin{figure}[!ht]
     \centering
     \includegraphics[width=0.5\textwidth]{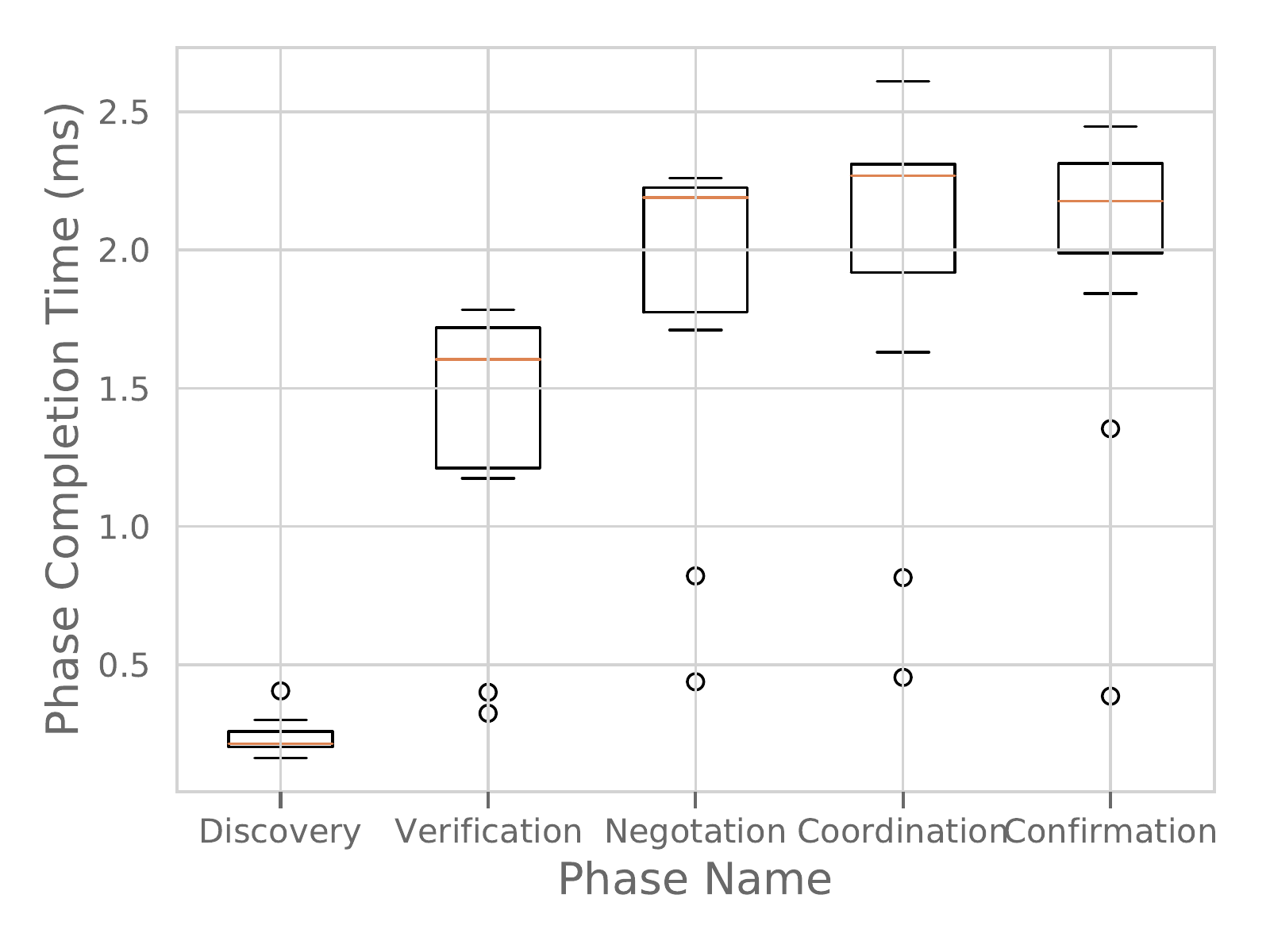}
     \caption{Average phase completion time in optimal V2V-CC environment with increasing concurrent consumers as variable. Lowest outliers for each phase are from the minimal number of concurrent consumers. The total completion time is total of all the phases.}
     \label{phaseWhisker}
 \end{figure}

Table \ref{tab:simulation_parameters} summarizes the parameters we used for the simulation. The number of 
suppliers varied between 1-10. In our protocol, an 
EV tries to connect to $1-3$ 
suppliers to obtain charging information, so $10$ 
suppliers provide a reasonable number of choices. The number of consumers is between $1$ and $21$. As the result section will show, increasing the number of consumers does not  increase the coordination time. Since the packets are small, we set the bandwidth to $24$ Mbps. We never fully utilized the available bandwidth. Therefore, allocating additional bandwidth does not improve the performance. For NDN experiments, we utilize $30$ and $50$ ms timeouts. That way, if a collision happens, the client quickly re-transmits the request. For mobility, we utilized $10, 30, 50$, and $70$ MPH, speeds that are close to actual speed on various types of road. Finally, for the IP experiments, we utilize $25$ms, $50$ms, and $100$ms delays between the client and the central controller. The delays are taken from surveys that point out the delay to current commercial cloud platforms\cite{shannigrahi2020next}.
\section{Results}

In this section, we first establish a baseline of the V2V-CC performance. We then investigate 
the behavior of the proposed protocol as the number of consumers increases.  Once the baseline is established, we 
compare the proposed protocol  
with a centralized (IP-based) approach where the 
EVs 
communicate with a central controller to find potential seller(s).


Figure \ref{phaseWhisker} shows the completion time of each 
protocol phase. Note that we could combine some of the phases but kept them separate for simplicity. 
In this experiment, we do not use any artificial losses, we consider minimal mobility speed, and we adopt an optimal consumer to 
supplier ratio of one supplier for every three consumers. The consumer also discovers a single 
supplier. As Figure. \ref{phaseWhisker} shows, each phase is complete within less than $2$ ms on average. With a low number of concurrent consumers, the completion times are reduced to below $0.5$ ms. 
Since NDN allows packet reuse through in-network caching, V2V-CC is able to reuse data packets both in the discovery and verification phases. The negotiation, coordination, and confirmation 
phases take longer times since they need to happen directly between the consumer and the seller. Even then, the average time for each of these phases is less than $2$ ms.



\begin{figure}[!h]
     \centering
     \includegraphics[width=0.45\textwidth]{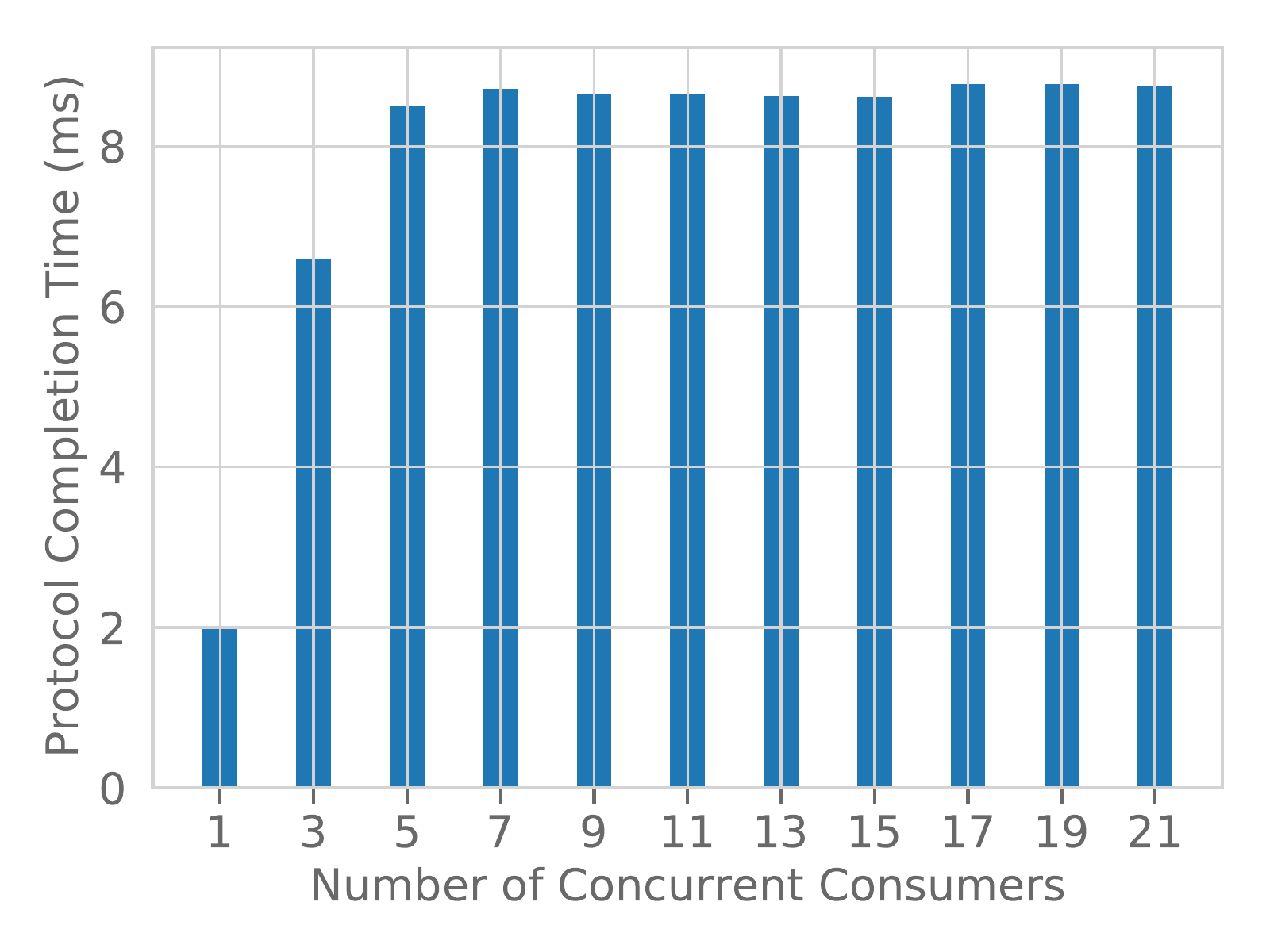}
     \caption{V2V-CC completion time in optimal environment with increasing number of concurrent consumers.}
     \label{concurrent_users}
 \end{figure}

We then tested V2V-CC with an increasing number of concurrent consumers. Figure \ref{concurrent_users} shows that 
as the number of consumers increases, the protocol completion time increases.
However, V2V-CC scales well beyond having five concurrent 
consumers with an optimal consumer to 
supplier ratio of three consumers to one 
supplier, as the completion time remains unchanged.

\begin{figure}[!h]
     \centering
     \includegraphics[width=0.45\textwidth]{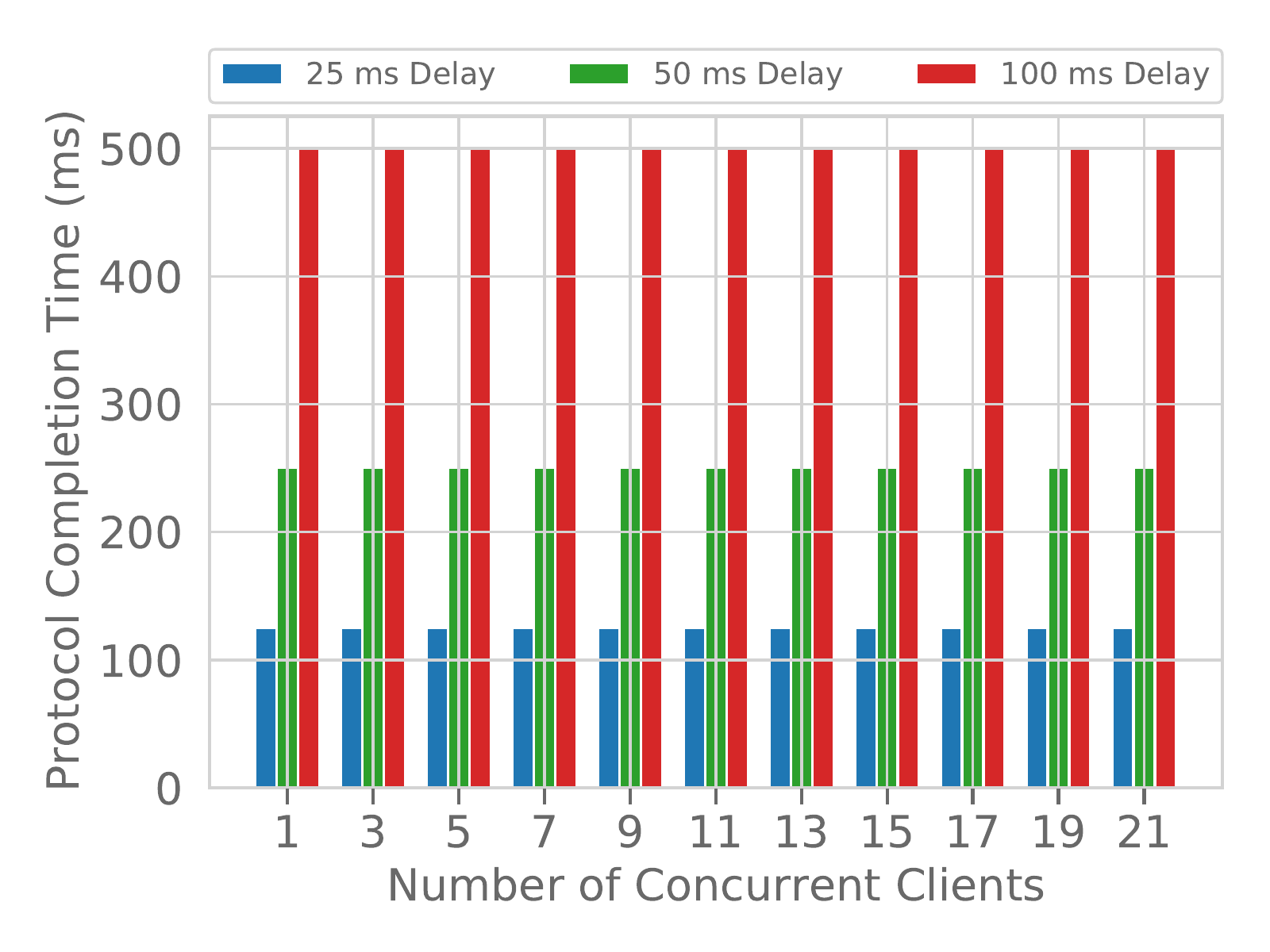}
     \caption{Three tests of IP-based central coordination. Each test is run with differing latency between the client and central coordinator.}
     \label{ip_cases}
 \end{figure}

Once we establish the baselines, we compare our protocol with an 
IP based central coordination approach. 
Figure \ref{ip_cases} shows the base cases for a central coordinator. We run tests using three types of delays in reaching the central coordinator, namely, $25$, $50$, and $100$ ms, respectively. These values represent typical delays to cloud computing platforms that we observed in our previous work \cite{shannigrahi2020next}. Each client begins by connecting to the central coordinator hosted in the cloud server taking one and a half round trip time (TCP handshake) and use a single packet to request the data that the server holds taking an additional RTT. Note that this time does not include any processing time at the central coordinator, which is likely to be an additional several hundred milliseconds.  In the best case (with $25$ ms one way delay), this approach takes over $125$ ms and in the worst case (with $100$ ms delay), it takes around $500$ ms.


  \begin{figure}
     \centering
     \includegraphics[width=0.45\textwidth]{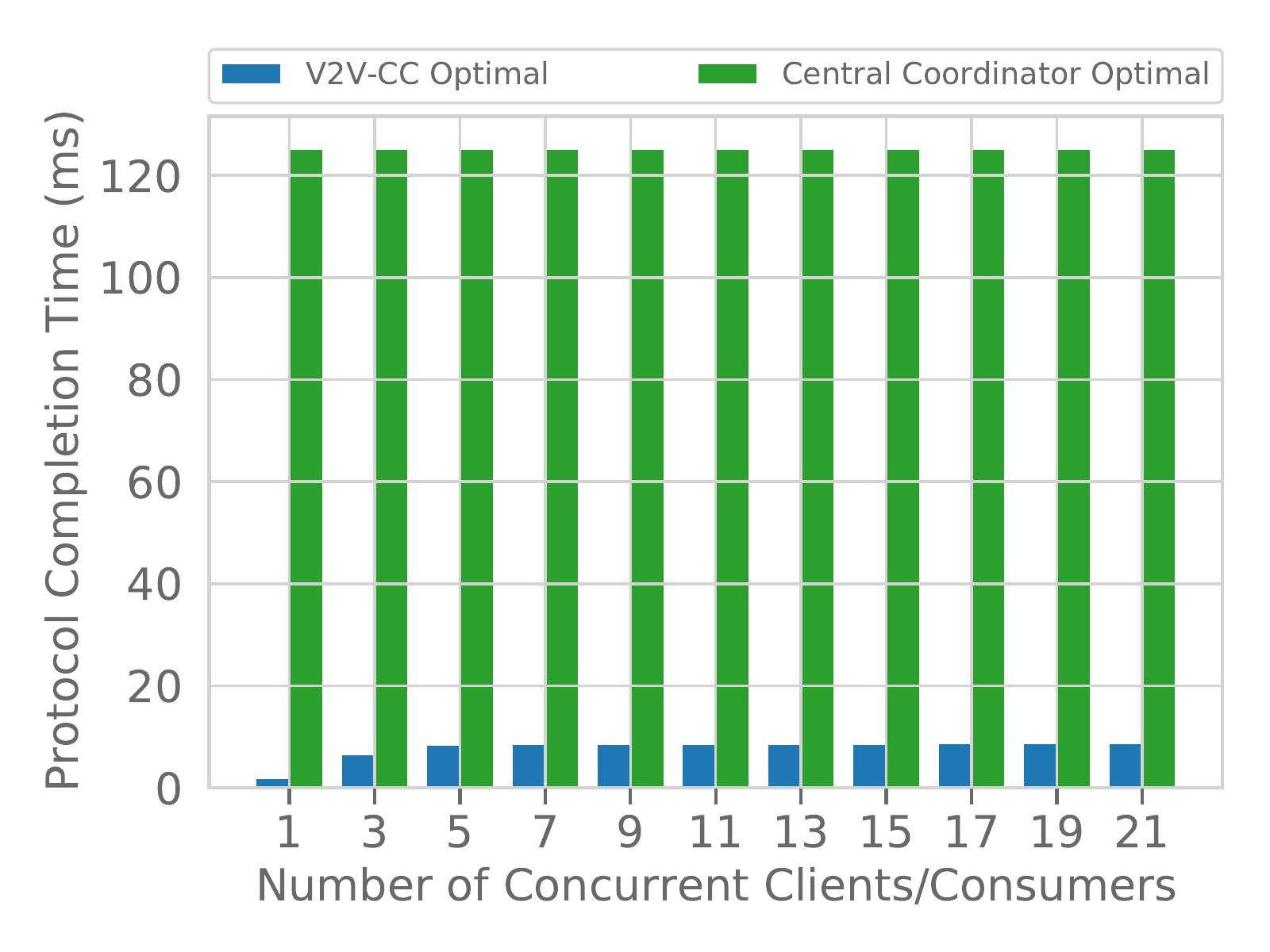}
     \caption{V2V-CC completion time in optimal environment compared with central coordination base case.}
     \label{v2v-vs-central}
 \end{figure}
 
In comparing V2V-CC with the IP-based approach, we observe that  even in the best case for the central coordination server, the additional latency of going to the cloud requires at least $125$ ms, while V2V-CC requires $8.79$ ms for 21 concurrent consumers, a $93\%$ reduction in time to complete charging coordination as Figure \ref{v2v-vs-central} shows. The reduction in latency is due to the peer-to-peer nature of V2V-CC since the distance between the consumer and the seller is minimal.

\begin{figure}
     \centering
     \includegraphics[width=0.45\textwidth]{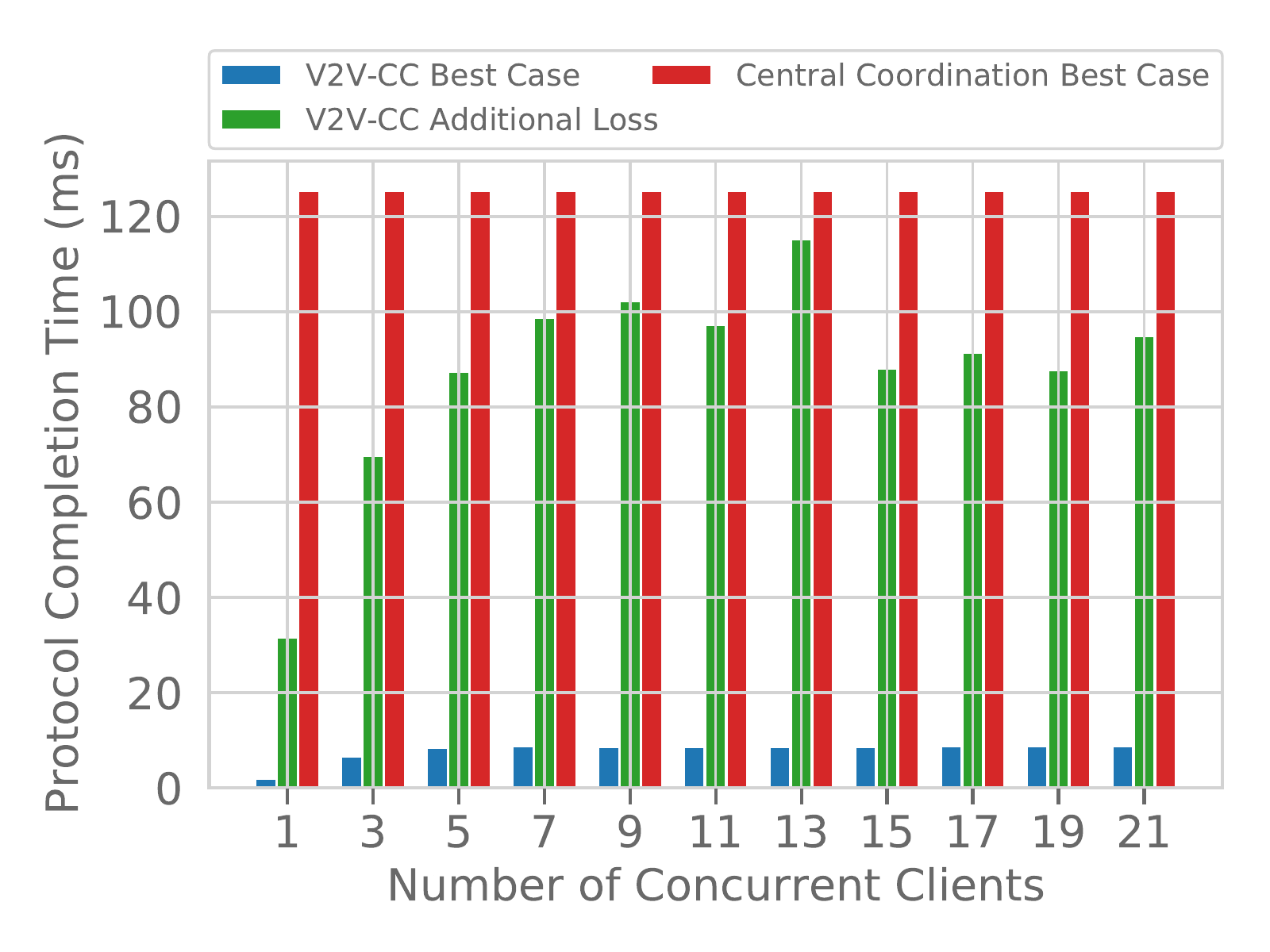}
     \caption{V2V-CC completion time in optimal environment and V2V-CC in an environment with $20\%$ additional loss. This data is an average of $10$ runs. This graph also includes the central coordination's best case (with $25$ ms delay and no losses).}
     \label{lossBar}
 \end{figure}

Even in the case of very high loss, V2V-CC remains scalable as shown in Figure \ref{lossBar}. Each set of loss tests were run with an added $20\%$ of artificial losses. 
Figure \ref{lossBar} shows that even with an extremely high packet loss rate, V2V-CC preforms similarly to a central coordinator that is working under ideal conditions (minimal delay and no losses). This is due to the fact that NDN uses in-network caching, which helps with fast retransmissions after packet loss. Additionally, serving content from cache reduces network congestion and also aggregates (using multicast and Interest aggregation) duplicate requests. Figure \ref{lossBar} shows that even with 20\% loss, V2V-CC works as well as the central controller's best case (25ms latency and no loss). It is well documented that any loss in TCP/IP will severely increase the total delay. In those scenarios, V2V-CC will further outperform the central coordinator approach. However, we omit those results for brevity.


 
\section{Discussions and Future Work}

While this work presents a novel V2V charging coordination mechanism, this is a preliminary study. For simulations, we have used WiFi for mobile connectivity. In real-world scenarios, the 
EVs are going to be connected either using 4G-LTE or 5G networks. However, 5G protocols are not fully standardized yet. Furthermore, the 4G-LTE stack was developed to work with IP, and therefore is not available for the NDN stack. 

While we have used WiFi, we have paid careful attention to make the simulation parameters (bandwidth and delay) to be consistent with the current 4G-LTE deployments. Note that utilizing WiFi does not reduce the generality of our study. For example, the 5G networks will reduce the last hop latency but the IP's point-to-point connection model will not change. An 
EV will still need to 
communicate with a central coordinator where the communication will be subjected to a number of variables (congestion, loss, delay, etc.). On the other hand, NDN will be able to utilize the 5G connectivity at Layer2 in the future. Finally, the goal of our study is not to compare IP's performance with that of NDN. We aim to demonstrate that a peer-to-peer V2V charging model is much faster than carrying out the coordination 
through a central controller. In the future, we plan to perform an emulation study where we evaluate V2V-CC over a 5G mobile network.



\section{Conclusion}
In this 
paper, we propose V2V-CC, a peer-to-peer charging coordination protocol over 
ICN. The V2V-CC uses named services to facilitate communications between the EVs that might be interested in selling and buying 
charging energy. Using extensive simulation studies, we show that V2V-CC is extremely fast and the whole protocol takes less than $10$ ms. We also compared V2V-CC with a centralized approach and we found that V2V-CC is orders of magnitude faster than an approach based on a central controller. Finally, even with extremely high losses (as high as $20\%$), V2V-CC is able to maintain similar level of performance to a central controller working under ideal conditions (with minimal delay and no losses). While our work is preliminary, we demonstrate that V2V charging coordination over ICN can pave the path for a faster, simpler, more flexible, and open charging marketplace where any EV can offer 
energy to 
interested EVs in a decentralized way.

\bibliographystyle{IEEEtran}  
\bibliography{references}

\end{document}